\documentstyle[aps,prl,epsfig,rotating,multicol]{revtex}

\begin{document}

\draft

\title{
Polarization and Strong Infra-Red Activity in Compressed Solid Hydrogen}

\author{Ivo Souza and Richard M. Martin}

\address{Department of Physics and Materials Research
Laboratory, University of Illinois, Urbana, Illinois 61801}

\date{\today}

\maketitle

\begin{abstract}
Under a pressure of $\simeq150$~GPa solid molecular hydrogen undergoes a phase
transition accompanied by a dramatic rise in infra-red absorption in the
vibron frequency range. We use the Berry's phase approach to calculate the
electric polarization in several candidate structures finding large,
anisotropic dynamic charges and strongly IR-active vibron modes. The
polarization is shown to be greatly affected by the overlap between the
molecules in the crystal, so that the commonly used Clausius-Mossotti
description in terms of polarizable, non-overlapping molecular
charge densities is inadequate already at low pressures and even more so for
the compressed solid.
\end{abstract}


\begin{multicols}{2}

\columnseprule 0pt

\narrowtext

The stretching mode of an isolated hydrogen molecule does not absorb infra-red
(IR) radiation, since the electric dipole moment vanishes by symmetry and
remains zero upon stretching. In the molecular solid the symmetry is lower, and
for some structures first-order IR activity of vibrational modes is 
allowed\cite{Zallen}. Indeed, some IR absorption occurs in the 
{\it broken-symmetry phase} (BSP or phase II)
\cite{footnote0}. More surprising is the dramatic increase in the
intensity of
IR absorption which occurs when crossing the boundary between the BSP phase and
 phase III (H-A) at around 150~GPa\cite{Hanfland93,Cui95,Mao94}. At 85~K the
vibron oscillator strength goes from $f \simeq 0.0002$ at 140~GPa to
$f \simeq 0.0134$ at 167~GPa\cite{Hemley97}.

A considerable effort has been made to try to explain this remarkable
behaviour. Hemley and co-workers\cite{Hemley} proposed that the increase in
molecular overlap leads in phase III to the formation of charge-transfer
states between neighboring molecules, with the coupling between a vibron and
charge-transfer excitations giving rise to charge oscillations between the 
molecules.
Baranowski\cite{Bar} invoked a strong ionization of the molecules in phase III.
Mazin {\it el al}\cite{Mazin} neglected the molecular overlap  
and treated the molecules as point objects which become
polarized in the quadrupolar field of the other molecules (EQQ model
\cite{Schnepp}). Edwards and Ashcroft\cite{Edwards} suggested that a
dielectric
instability associated with a charge-density wave causes an enhancement of the
molecular dipole moments in phase III. Some theories\cite{Bar} focus
on purely static charge, others\cite{Hemley} on purely dynamic (i.e.,
displacement-induced\cite{Harrison}) and the others
\cite{Mazin,Edwards} take the two contributions to be comparable.
There is also no agreement on the relative importance of intramolecular
\cite{Bar,Mazin,Edwards} and intermolecular\cite{Hemley} charge transfer
\cite{Hemley97}. To settle these issues, a careful treatment of the bulk
electric polarization, which is the central quantity at play in the physics of
IR absorption, is needed.

For that purpose we have used the ``modern theory of polarization''
\cite{Vanderbilt}, which shows how to compute
the polarization 
of a periodic insulating system as a Berry's phase
derived from the electronic wave functions. The
information about the polarization is
therefore in the {\it phases} of the wave functions, not in the periodic
density alone, whose dipole moment is in general ill-defined,
depending on the choice of unit cell\cite{RMM}.
The density-based picture is valid only in the
Clausius-Mossotti limit of localized, polarizable units with
non-overlapping charge densities\cite{Szigeti} and breaks down whenever
significant charge delocalization
prevents an unambiguous assignment of
the electron density to particular atoms, molecules or ions in the solid.
Such overlap effects are known to play an important role
in the absorption of light by lattice vibrations in ionic systems
\cite{Szigeti}, which typically have a much larger overlap than molecular
crystals\cite{Harrison}. But at 150~GPa solid hydrogen has been compressed by
almost a factor of ten\cite{Loubeyre},
hence the motivation for applying the modern theory of polarization to this
problem.

Using the Berry's phase approach we have calculated the Born
effective charge tensors $Z^*$\cite{Harrison,Bruesch}. In order to compare with
experiments done on polycrystalline samples, we used the expression for
the oscillator strength $f(j)$ of the contribution of a TO phonon mode {\it j}
to the dielectric function averaged over all directions\cite{Bruesch},

$$f(j)={ {4\pi} \over {\Omega \omega(j)^2}}\left( {1 \over 3}
\sum_{\alpha=1}^3 \xi_{\alpha}(j)^2\right),$$

\noindent where
$\xi_{\alpha}(j) \equiv {1 \over \sqrt{m_{\rm p}}}\sum_{\rm k}\sum_{\beta}
Z^{*}_{\alpha \beta}\left( k \right) e_{\beta}\left( k;j \right)$,
$k$ indexes
the atoms in the unit cell of volume {\it $\Omega$},
$e_{\beta}\left( k;j \right)$ are the normalized
eigenvectors associated with the mode, $\omega(j)$ is its angular
frequency and $m_{\rm p}$ is the proton mass.

The crystal structure in the high-pressure phases
has not yet been determined experimentally\cite{Loubeyre}. For clarity we have
investigated two of the simplest candidate structures for the
compressed solid which have IR-active vibron modes\cite{Cui95} and are also
energetically favorable\cite{Kaxiras,Natoli95}.
The parameters $a$ (lattice constant), $c/a$, bond length and $\theta$ of
$Cmc2_1$ and $C2/m$ (see Fig. 1) were
optimized by minimizing the enthalpy at fixed pressure using a variable cell
shape method\cite{Souza}, and they are collected in Table I for several
densities\cite{note_rs}. In some cases the centers of the molecules were
allowed to move away from the {\it hcp} sites by symmetrical amounts in the
{\it yz} plane, which does not lower the symmetry while allowing for
the manifestation of the instability reported in Ref.\cite{Edwards}. The
calculations were done in the local density approximation (LDA) to the
density functional theory\cite{noteLDA}, with clamped nuclei.

\epsfig{file=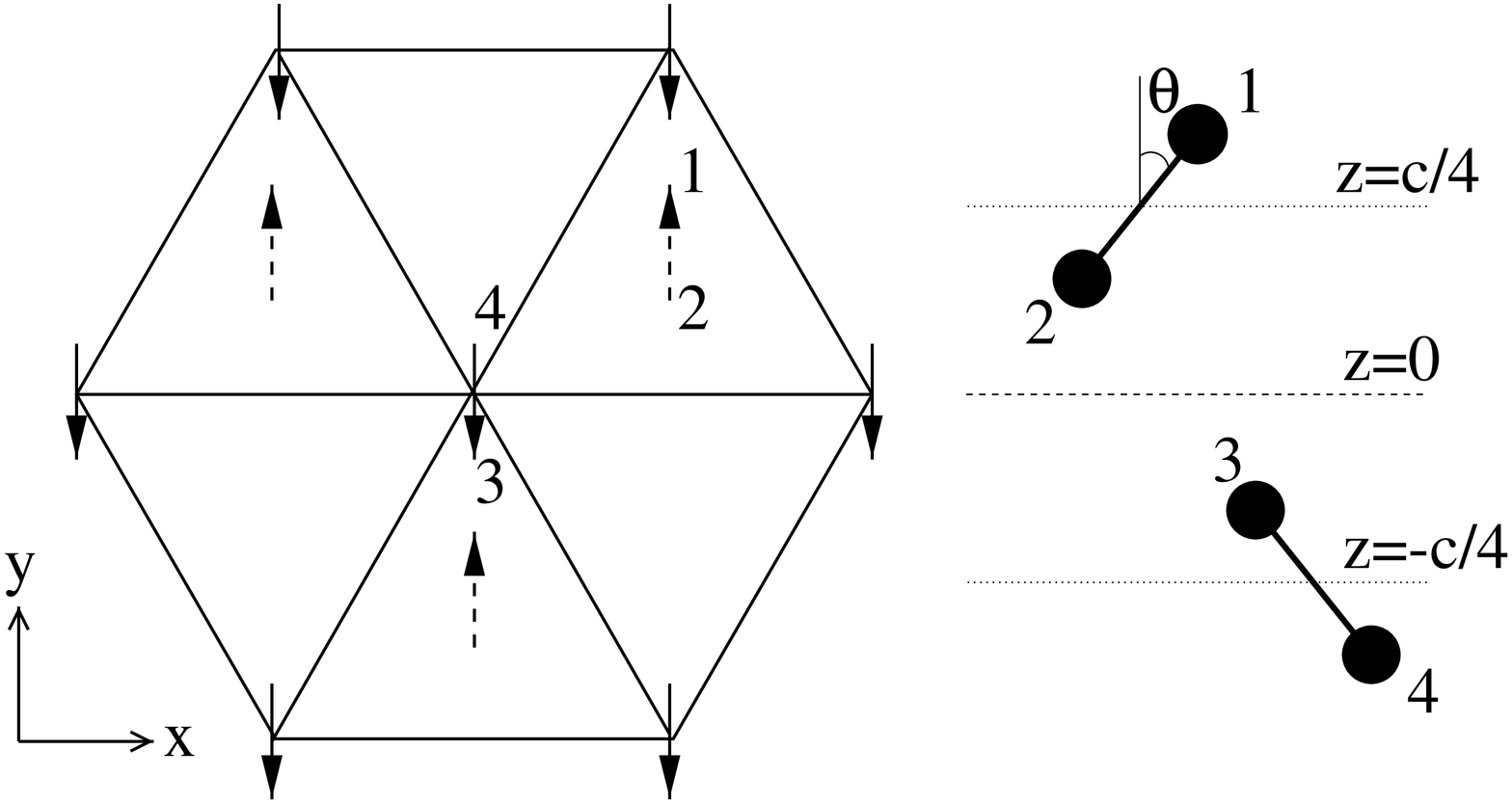,width=1.45in,angle=-90}

\begin{minipage}{3.4in}
{\small
FIG. 1. The $Cmc2_1$ structure viewed along the c-axis (left) and in the
$yz$ plane (right). The centers of the molecules lie on {\it hcp} sites, and
the molecules in the two sublattices are tilted away from the c-axis
by opposite angles $\theta$ and $-\theta$. The $C2/m$ structure is identical
except that the two molecules are tilted in the same direction by an angle
$\theta$.}
\end{minipage}\\ \\

\begin{tabular}{cccccccc}
Structure & P & $r_{\rm s}$ & c/a & $r_{\rm bond}$ &
$\theta$ & $f_{\rm in-phase}$ & $f_{\rm out-phase}$\\
\hline
\hline
$C2/m$ & 115 & 1.52 & 1.588 & 1.456 & 69.5 & 0 & 0.169\\
$C2/m^{\dag}$ & 115 & 1.52 & 1.583 & 1.460 & 70.6 & 0 & 0.589\\
\hline
$Cmc2_1$ & 13 & 2.0 & 1.576 & 1.445 & 54.0 & $10^{-6}$ &
$9\times10^{-6}$\\
$Cmc2_1$ & 115 & 1.52 & 1.576 & 1.445 & 54.0 & 0.00079 & 0.0114\\
$Cmc2_1^{\dag}$ & 115 & 1.52 & 1.574 & 1.451 & 56.6 & 0.014 & 0.049\\
$Cmc2_1$ & 152 & 1.47 & 1.571 & 1.443 & 55.2 & 0.0026 & 0.026\\
$Cmc2_1$ & 167 & 1.45 & 1.570 & 1.443 & 55.6 & 0.004 & 0.034\\
$Cmc2_1$ & 180 & 1.43 & 1.569 & 1.443 & 55.9 & 0.0059 & 0.044\\
\hline
Phase III & 167 & & & & & undetected & 0.0134\\
\hline
\end{tabular}\\
\vspace{0.22cm}
\begin{minipage}{3.4in}
{\small
$^{\dag}$ Centers of the molecules are off-site (see text). The off-site
displacement in a.u. of the upper molecule in Fig. 1 is
$\delta y = -0.1$,
$\delta z = -0.01$ ($C2/m$), and $\delta y = -0.07$ ($Cmc2_1$).\\

TABLE I. Optimized parameters (except for $r_s=2.0$) and vibron oscillator
strengths for several pressures.
Pressures are in GPa and are obtained using the
equation of state of Ref.\cite{Loubeyre}, angles are in degrees,  and the
other quantities in atomic units (a.u.). The experimental results for phase III
are taken from Ref.\cite{Hemley97}.
``undetected'' means that the Raman-active vibron is not observed in the IR 
spectra.
}
\end{minipage}\\

The existence of a center of inversion between the two molecules in the
unit cell of $C2/m$ leads to the relation $Z^*(1)=-Z^*(2)=-Z^*(3)=Z^*(4)$
between the effective charge tensors of the four atoms. For the optimized
structure at 115~GPa we obtained, for atom 2 before and after the relaxation
away from the hcp sites, respectively:\\

{\small $Z^{*}\simeq\left( \begin{array}{ccc}
<0.005 & 0 & 0\\
0 & 0.53 & 0.18\\
0 & 0.08 & 0.07
\end{array} \right)$},
{\small $Z^{*}\simeq\left( \begin{array}{ccc}
0.05 & 0 & 0\\
0 & 0.96 & 0.28\\
0 & 0.15 & 0.16
\end{array} \right)$}.\\

Although $Z^*_{xx}$ is non-zero by symmetry, we found it to be very small
before the off-site relaxation.
The pronounced anisotropy of $Z^*$ implies large dynamic charges,
since displacements of rigid ions (static charges) give rise to isotropic
diagonal $Z^*$ tensors\cite{Harrison}.
The smallness of $Z^*_{xx}$ relative to the components in the
{\it yz} submatrix suggests that
most of the charge transfer occurs in the planes of the molecules, since a
displacement $\delta x$ of an atom in the {\it x} direction changes the
distances to other atoms in that plane to second order only. This analysis
establishes that intramolecular and/or in-plane intermolecular charge transfer
are the dominant effects, but it does not distinguish between the two
\cite{note_current}.
The small displacement
away from the {\it hcp} sites has a large effect on $Z^*$, revealing a strongly
nonlinear dependence of the polarization on the positions of the molecules.

Such large effective charges give rise to very strong IR absorption by the
vibron mode in which the two molecules in the primitive cell vibrate
out-of-phase (the in-phase mode is Raman active but IR-inactive in this
structure\cite{Cui95}).
Before the relaxation, the oscillator strength at 115~GPa is about 13 times
larger than the experimentally measured value in phase III at 167~GPa, and it
increases by a factor of 3.5 upon relaxation (see Table I). Due to closure of
the LDA band gap, this structure was not studied at significantly higher
pressures.

In $Cmc2_1$ the symmetry of the structure dictates the following form
for the effective charge tensors:\\

{\small $Z^{*}(1)=\left( \begin{array}{ccc}
a & 0 & 0\\
0 & b & c\\
0 & d & e
\end{array} \right)$},
{\small $Z^{*}(2)=\left( \begin{array}{ccc}
-a & 0 & 0\\
0 & -b & c'\\
0 & d' & -e
\end{array} \right)$},\\
{\small $Z^{*}(3)=\left( \begin{array}{ccc}
a & 0 & 0\\
0 & b & -c\\
0 & -d & e
\end{array} \right)$},
{\small $Z^{*}(4)=\left( \begin{array}{ccc}
-a & 0 & 0\\
0 & -b & -c'\\
0 & -d' & -e
\end{array} \right)$}.\\

\begin{tabular}{cccccccc}
$r_{\rm s}$ & a & b & c & d & e & c' & d' \\
\hline
\hline
2.0  & -0.0178 & -0.0033 & 0.028 & 0.044 & 0.0005 &
0.042 & 0.040\\
1.52 & -0.022 & -0.14 & 0.069 & 0.193 & -0.0088 & 0.208 & 0.271\\
$1.52^{\dag}$ & -0.050 & -0.30 & -0.014 & 0.091 & -0.0087 & 0.243 & 0.348\\
1.47 & -0.020 & -0.20 & 0.069 & 0.216 & -0.015  & 0.260 & 0.347\\
1.45 & -0.018 & -0.229 & 0.068 & 0.223 & -0.017 & 0.283 & 0.380\\
1.43 & -0.017 & -0.254 & 0.065 & 0.228 & -0.020 & 0.307 & 0.413\\
\hline
\end{tabular}\\
{\small  $^{\dag}$ Centers of the molecules are off-site (see text).

TABLE II. Effective charges in atomic units for the optimized $Cmc2_1$
at several densities.
}\\

The calculated values for several densities are shown in Table II.
The anisotropy is again very large, which implies large dynamic charges, but
the overall values are somewhat smaller than for $C2/m$.
Both vibrons are IR-active in this structure\cite{Cui95}:
the out-of-phase vibron
has a large oscillator strength in
reasonable agreement with the measured values in phase III (see Table I),
whereas the in-phase vibron is about an order of magnitude weaker,
which would still be enough to
be detected experimentally, if $Cmc2_1$ was the correct structure for that
phase.
The intensity of the absorption grows rapidly with pressure, in agreement with
the behavior in phase III\cite{Cui95,Hemley97,Hemley}.
At 167~GPa we obtain for the Szigeti effective charge
$q^*$\cite{Hemley97,Szigeti} of the out-of-phase vibron $q^* \simeq 0.047e$,
comparable to the measured value of $q^* \simeq 0.032e$\cite{Hemley97}, and its
pressure derivative is $d q^* /dP \simeq 2.3 \times 10^{-4}$ {\it e}/GPa,
close to the experimental value of
$\simeq 2.1 \times 10^{-4}$ {\it e}/GPa\cite{Hemley97}.
The effect of off-site relaxation of the molecules is again significant, with
the oscillator strengths of the vibrons increasing 18-fold (in-phase) and
4-fold (out-of-phase) at 115~GPa.

\epsfig{file=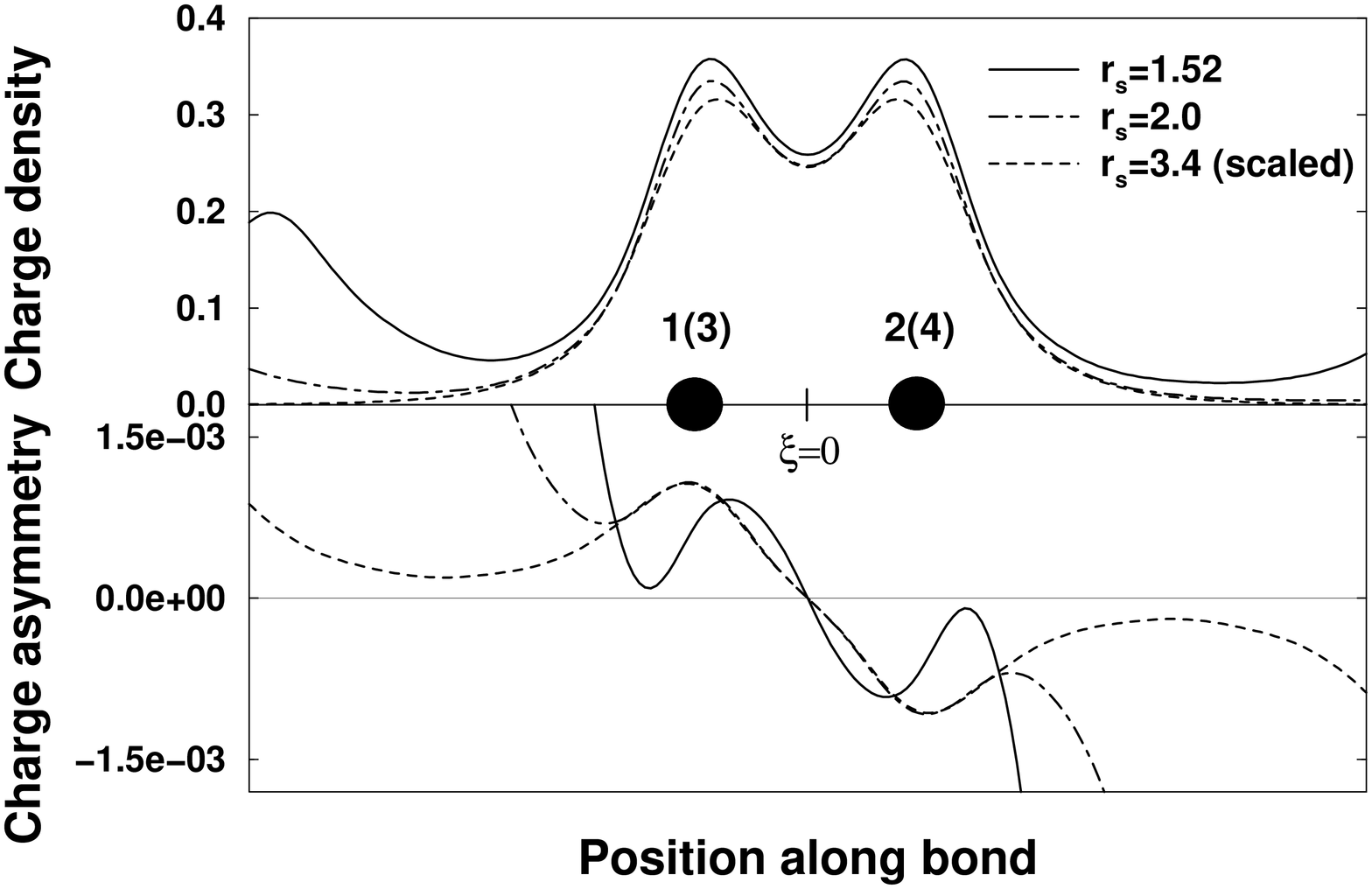,width=2.2in,angle=-90}\\ \\
\begin{minipage}{3.4in}
{\small
FIG. 2. (a) electron density
$n({\bf r})$ in a.u. for $Cmc2_1$ along a line
joining atoms 1(3) and 2(4) of Fig. 1. (b) asymmetry along that
line, defined as $a(\xi)=\left[n(\xi)-n(-\xi)\right]/2 n_0$, where $n_0$ is
the maximum electronic density in the hydrogen atom\cite{Edwards}. For
$r_s=3.4$ $a(\xi)$ was
multiplied by $\left( 3.4/2.0 \right)^4$ for comparison with $r_s=2.0$.
The molecular centers are on {\it hcp} sites, and therefore the intramolecular
$a(\xi)$ is very small\cite{Edwards}, and yet the vibron oscillator strength
for $r_s=1.52$ is quite large.
}
\end{minipage}\\ \\

More insight into the physics of polarization can be obtained by looking at the
electron density.
In Fig. $2a$ we show the density along the axis of a
molecule in the $Cmc2_1$ structure for several pressures. The asymmetry induced
by the neighboring molecules can be clearly seen in the higher pressure curve,
and is more visible in Fig. $2b$, where we have plotted the asymmetry function
$a(\xi)$\cite{Edwards}.  If the molecules can be considered to be
separate polarizable objects, as in the Clausius-Mossotti model,
then the asymmetry of the molecular density will determine the
polarization.  For example, in the widely used EQQ model
\cite{Mazin,Schnepp,Silvera80,Kra}
the molecular dipoles are induced by the electric field at the
centers of the molecules due to the quadrupole moments $Q$ of the other
molecules. In this model the intramolecular asymmetry scales as $1/r_s^4$ 
(assuming that $Q$ and the molecular polarizability do
not change much with pressure\cite{Evans}).
Indeed this is the case at low and moderate pressures ( $r_s \gtrsim 2.0$),
as can be seen in Fig. $2b$ from the agreement in the intramolecular region
between the curve for $r_s=2.0$ and the curve for $r_s=3.4$ scaled by
$\left( 3.4/2.0 \right)^4$. However, the scaling clearly does not hold up to 
the megabar
range, since the intramolecular $a(\xi)$ would have to be around 3 times
larger for $r_s=1.52$ than for $r_s=2.0$, whereas it is found to be
slightly smaller and decreasing with pressure in that range. Therefore in
the megabar range the EQQ interactions do not account for the local
field acting on the molecules, contrary to the model used in
Ref.\cite{Mazin}.

In Ref.\cite{Edwards} $a(\xi)$ in the intramolecular region 
was interpreted as
an estimate of the dipole moment of the molecule in the solid, which leads to
the prediction that the vibron oscillator strength should scale roughly as
the square of its magnitude. Instead, we have found that although the
intramolecular asymmetry is 
comparable
for $r_s=1.52$ 
and
$r_s=2.0$,
the oscillator strengths of the 
vibrons are $\sim 800$ (in-phase) and $\sim 1300$ (out-of-phase) times larger 
for $r_s=1.52$
(see Table I).
The origin of this surprising result can already be seen in Fig. 2, where it is
clear that for both these densities $a(\xi)$ is
large 
in the intermolecular region, where the density is small, which is related to
the fact that the outer regions of the molecules are most easily 
polarizable\cite{Werner}. Thus, 
when considering polarization effects, 
molecular overlaps
may be important, even if they are small, causing
the dipole moments of the molecules 
to be
ill-defined
and compromising the Clausius-Mossotti picture. 

At low enough densities the Claussius-Mossotti model must be correct; however,
to our knowledge there has never been a thorough investigation of
{\it how low the density must be for this limit to apply}.
This is now possible using the modern theory of polarization\cite{Vanderbilt}.
Here we present results for the $Cmc2_1$ structure as a function of volume.
(We emphasize that this is not meant to represent real hydrogen at low 
pressure, where the molecules are not oriented but instead behave as quantum
rotors\cite{Silvera80,Kra}, but is sufficient to establish the desired points
regarding the sensitivity of the polarization to the molecular overlap.)
In this structure there is a net spontaneous polarization
{\bf $\rm P_0$} along the {\it c}-axis, and in Fig. 3 we plot {\bf $\rm P_0$}
as a function of $r_s$, together with the value given by the EQQ
model and the dipole moment for a particular choice of unit cell.
The cell we used goes from $z=0$ to $z=c$ in Fig. 1, which is the
obvious choice in the low pressure limit since the electronic density
will vanish at the boundaries and the dipoles will become well-defined
in that limit. For $r_s \geq 1.52$
atoms 1 and 3 in Fig. 1 have an excess of electron charge with respect to atoms
2 and 4, as inferred from $a(\xi)$. In a Clausius-Mossotti framework this
would lead to a net
polarization (sum of the molecular dipole moments) along minus
{\it z}, and that is indeed the result of the EQQ model (and also of the 
dipole moment for $r_s > 2.5$), but 
between $r_s \simeq 1.7$ ($\simeq 50$~GPa) and
$r_s \simeq 3.0$ (around atmospheric pressure)
the Berry's phase polarization has the opposite sign.
Most surprising is the fact that the corrections to the models are
so large even at low pressures (large $r_s$).
Since the three quantities in Fig. 3 start to converge to each other only for 
$r_s \gtrsim 3.0$, we conclude that
the Clausius-Mossotti limit is reached only at {\it negative}
pressures. A possible explanation for this observation is that
at zero pressure there must still be some overlap of the
wavefunctions in order to balance the attractive interactions, 
and only at negative pressure does this overlap become truly negligible.

\epsfig{file=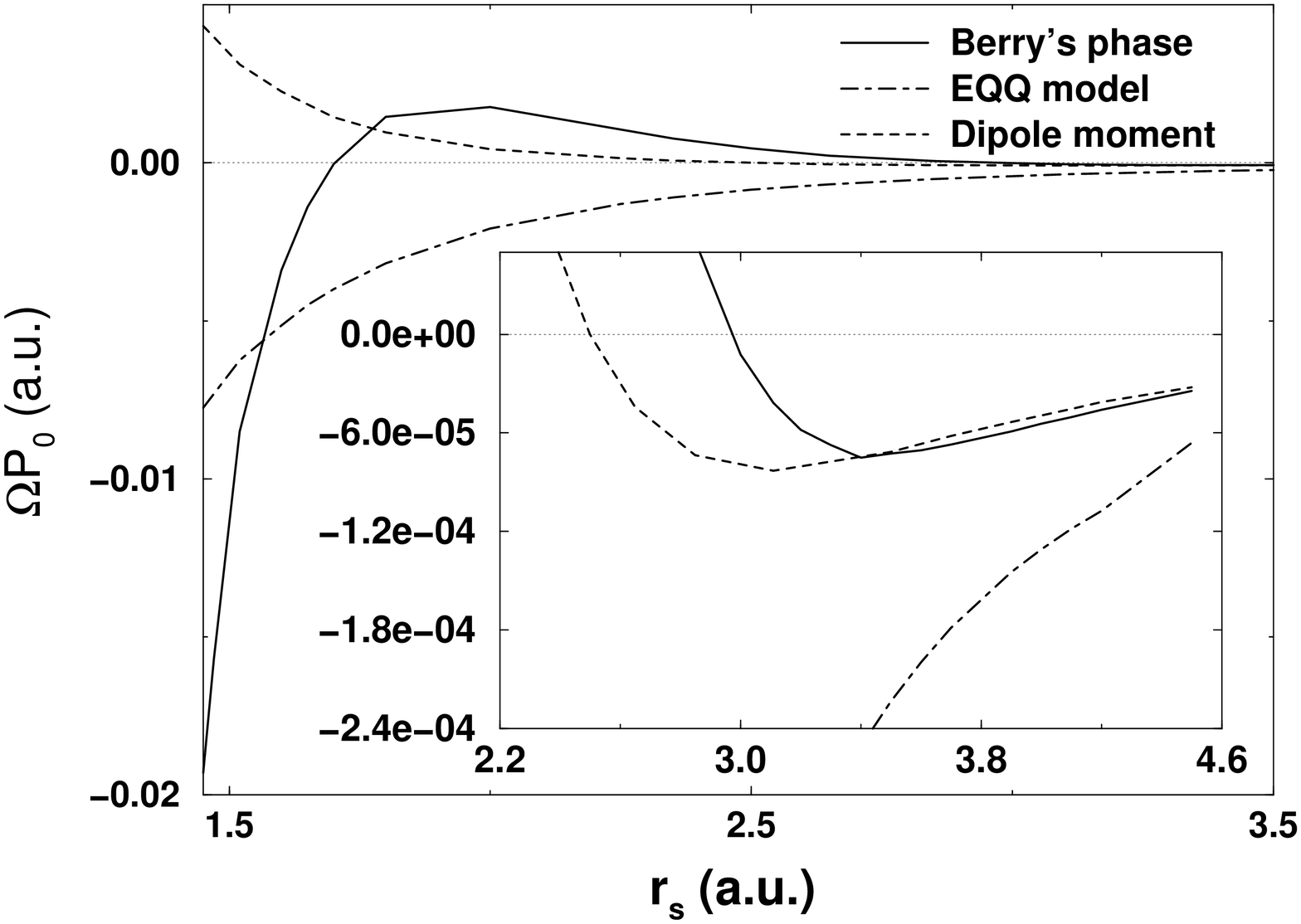,width=2.3in,angle=-90}\\
\begin{minipage}{3.4in}
{\small
FIG. 3. Spontaneous polarization per unit cell as a function of $r_s$
for the {\it hcp}-centered $Cmc2_1$.
The structural
parameters are the optimized parameters for $r_s = 1.52$ (see Table I).
The Berry's phase result is compared with the value given by
the EQQ model, using the quadrupole moment and
polarizability of the H$_2$ molecule quoted in Ref.\cite{Kra},
and with the ``dipole moment'' (see text) of the charge density in a unit 
cell. 
}
\end{minipage}\\ 

In summary, we have presented the first {\it ab-initio} calculations of the
intensity
of IR absorption in solid hydrogen, showing that the intense vibron IR activity
found in some 
structures is caused mainly by large dynamic charges.
The strong dependence of the effective charges on the symmetry and structural
parameters
offers a plausible explanation for the difference in IR
absorption between the two high-pressure phases, and the study of other
structures with the present approach may help identify the crystal
structure of those phases.
The physics of polarization of solid hydrogen 
appears to be more subtle and
less amenable to simple models than has been previously assumed. This seems to
be caused by the very delocalized nature of the induced dipoles, 
which extend out to the exponentially decaying
tails of the molecular charge density\cite{Werner},
rendering the bulk 
polarization very sensitive to even small overlaps between the molecules.
These findings may be relevant for other molecular solids, since the
Clausius-Mossotti picture is assumed in most treatments of the IR
absorption in such systems\cite{Califano}.

It is a pleasure to thank J. L. Martins for providing the 
LDA program and for many illuminating discussions.
We also benefited from conversations with N. W. Ashcroft, B. Edwards,
A. F. Goncharov, R. Resta, S. Scandolo, I. F. Silvera and
R. Zallen. The calculations were done in part at the NCSA.
This work was supported by the DOE Grant No. DEFG02-96-ER45439 and the
NSF grant No. DMR9422496.
I.S. 
acknowledges financial support from JNICT (Portugal) 
and  R.M.M. thanks the Aspen Center for Physics 
for its hospitality during part of this work.

\end{multicols}


\begin{references}
\bibitem{footnote0} In the low pressure phase
of para-hydrogen
IR activity is not allowed by symmetry, and the very weak absorption that is
observed is believed to be disorder-induced.
\bibitem{note_rs} The density is expressed in terms of $r_s$, defined in
atomic units as $4 \pi r_s^3 / 3 = V/N$, where $N$ is the number of atoms
in the volume $V$.
\bibitem{noteLDA} We used a plane-wave cutoff of 90~Ry and the bare
proton potential. The self-consistent calculations with 4 atoms per cell used a
(11,11,11)  Monkhorst-Pack mesh, and the Berry's phase calculations were done
with a (28,16,16) mesh
(fewer k-points were used at low pressure).
\bibitem{note_current} Preliminary results from a microscopic analysis of the
polarization current
associated with a vibron seem to indicate that both intramolecular and
intermolecular charge transfer remain important down to low pressures.
\end{references}
\end{document}